\documentclass[%
reprint,
showpacs,
preprintnumbers,
amsmath,
amssymb,
aps,
prl,
floatfix,
lengthcheck
]{revtex4-1}
\usepackage{graphicx}
\usepackage[hidelinks]{hyperref}
\usepackage{braket}
\usepackage{upgreek}
\usepackage{subscript}

\newcommand{\rubidium}{\textsuperscript{87}Rb}
\newcommand{\us}{$\upmu$s}
\newcommand{\um}{$\upmu$m}
\newcommand{\uK}{$\upmu$K}

\newcommand{\rbohr}{\textit{a}\textsubscript{0}}
\newcommand{\density}[2]{#1$\times$10\textsuperscript{#2}\unit{}{cm\textsuperscript{-3}}}
\newcommand{\vecx}{\vec{r}}

\newcommand{\RydGS}{\textit{nS}\,+\,5\textit{S}}
\newcommand{\RydSpecificGS}{40\textit{S}\,+\,5\textit{S}}
\newcommand{\RydNGS}{\textit{nS}\,+\,\textit{N}$\times$\,5\textit{S}}

\newcommand{\unit}[2]{\mbox{#1\,#2}} 
 
\newcommand{\trapfreq}[2]{$\omega$\textsubscript{#1}\,=\,2$\pi$$\times$\unit{#2}{Hz}}
\newcommand{\eRb}{\mbox{\textit{e}\textsuperscript{-}-Rb(5\textit{S})}}

\begin{document}

\title{Probing a scattering resonance in Rydberg molecules with a Bose-Einstein condensate}

\author{Michael~Schlagm\"{u}ller}
\author{Tara Cubel Liebisch}
\author{Huan Nguyen}
\author{Graham Lochead}
\author{Felix Engel}
\author{Fabian~B\"{o}ttcher}
\author{Karl M.\ Westphal}
\author{Kathrin S.\ Kleinbach}
\author{Robert L\"{o}w}
\author{Sebastian Hofferberth}
\author{Tilman Pfau}
\email{t.pfau@physik.uni-stuttgart.de}
\affiliation{5. Physikalisches Institut and Center for Integrated Quantum Science 
and Technology, Universit\"{a}t Stuttgart, Pfaffenwaldring 57, 70569 Stuttgart, 
Germany}
\author{Jes\'{u}s P\'{e}rez-R\'{i}os}
\author{Chris H.\ Greene}
\affiliation{Department of Physics and Astronomy,
Purdue University, 47907 West Lafayette, IN, USA}
\date{23 October 2015}

\begin{abstract}
We present spectroscopy of a single Rydberg atom excited within a Bose-Einstein condensate.  We not only observe the density shift as discovered by Amaldi and Segr\`{e} in 1934~\cite{Amaldi}, but a line shape which changes with the principal quantum number \textit{n}. The line broadening depends precisely on the interaction potential energy curves of the Rydberg electron with the neutral atom perturbers. In particular, we show the relevance of the triplet \textit{p}-wave shape resonance in the \eRb{} scattering, which significantly modifies the interaction  potential. With a peak density of \density{5.5}{14}, and therefore an inter-particle spacing of \unit{1300}{\rbohr} within a Bose-Einstein condensate, the potential energy curves can be probed at these Rydberg ion - neutral atom separations.
We present a simple microscopic model for the spectroscopic line shape by treating the atoms overlapped with the Rydberg orbit as zero-velocity, uncorrelated, point-like particles, with binding energies associated with their ion-neutral separation, and good agreement is found.
\end{abstract}

\maketitle

A Rydberg atom excited in a dense background gas of atoms provides a testbed of Rydberg electron-neutral atom collisions. Spectroscopy has always been a sensitive technique for studying these collisions and in particular spectroscopy performed in a cold, dense atom sample eliminates most other line broadening mechanisms, thereby isolating the effects of elastic and inelastic electron-neutral atom collisions on the line shape.  The realization of ultralong-range Rydberg molecules via elastic electron-neutral collisions~\cite{Greene2000} relies on cold, dense atom samples.  Since the first observation~\cite{Bendkowsky2010}, Rydberg molecules continue to be realized with increasing experimental control in various potential energy landscapes and atomic species~\cite{Bendkowsky2010,Gaj2015, Tallant2012, DeSalvo2015, Sassmannhausen2015, Anderson2014,Krupp2014, Bellos2013}, where the neutral atom ground state wavefunction is typically bound in the outer one or two potential wells of the electron-atom potential energy curves (PECs).  By applying an electric field~\cite{Li2011} or by exciting \textit{nS} states with nearly integer quantum defects, as in Cs~\cite{Shaffer2014}, more deeply bound trilobite Rydberg molecules~\cite{Greene2000} are realized.  With higher densities of cold atom samples, many neutral atoms overlap with the Rydberg orbit and the bound states become unresolvable~\cite{LossesNature, Gaj2014}.  By utilizing the high densities of a Bose-Einstein condensate (BEC), the neutral atoms within the Rydberg orbit provide a probe of elastic and inelastic electron-neutral collisions for a large range of ion-neutral separations.  We show in this paper that Rydberg spectroscopy in a BEC allows us to probe, with high resolution, the scattering resonance directly for the first time in this temperature regime.

In a completely different temperature regime (\unit{\textgreater\,400}{K}) than the work presented here (\unit{\textless\,1}{\uK}), Rydberg spectroscopy done during the 1980s in unpolarized thermal vapors investigated the line shift and line broadening of Rydberg atoms excited in a background gas of the same species atoms at similar densities~\cite{Weber1979, Stoicheff1980}.  Subsequent theory work~\cite{Fabrikant1986, Borodin1990, Borodin1991, Henry2002} modeled the line shapes by taking into account the scattering resonance and an interplay of elastic and inelastic collisions. 

In this work we present a comprehensive experimental and {\it ab initio} theoretical study of the nature of the line shape broadening associated with a Rydberg atom immersed in a BEC.  In particular it is found that elastic electron-neutral scattering near the crossing of the \textit{nS} and the butterfly shape resonance potential curves~\cite{Hamilton2002} dominate the line broadening behavior. The agreement between the simulated line shapes and spectra of the Rydberg BEC spectroscopy also demonstrates that this spectroscopy is a probe of the PECs over a large range of inter-nuclear separations, offering a tool for testing theoretical PECs.

\begin{figure}
\centering
	\includegraphics[scale = 1]{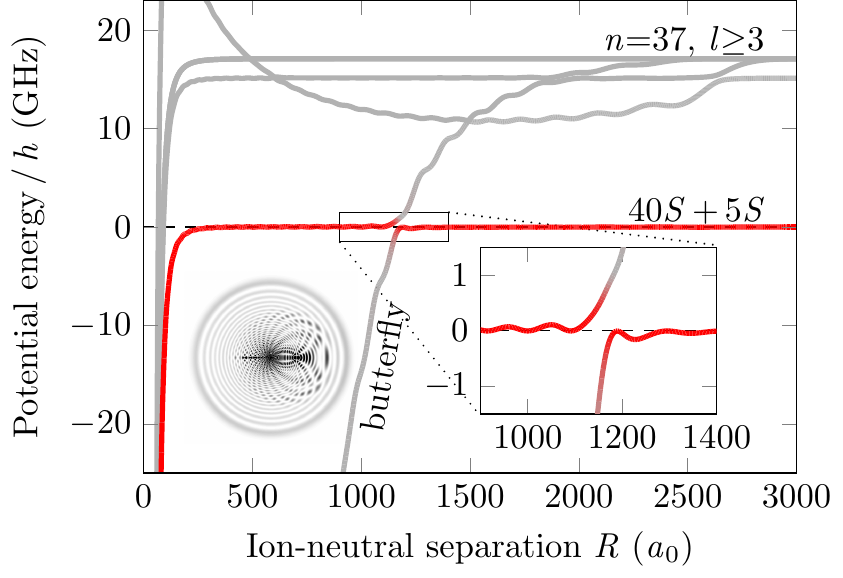}
	\caption{(Color online) Potential energy curves around the \RydSpecificGS{} state. The red color indicates the \textit{S}-character of the state and therefore the probability to excite into this state due to the selection rules. The butterfly shape resonance~\cite{Hamilton2002} has an anti-crossing with the \RydSpecificGS{} at around \unit{1200}{\rbohr}.  Neutral perturbers inside this region can shift the resonance frequency tens of MHz or more, therefore causing significant spectral broadening. The left inset shows a 2d cut trough the electron density of a butterfly state at \unit{1175}{\rbohr}. The right inset is a zoom in on the crossing of the shape resonance and the \RydSpecificGS{} state.}
	\label{fig:butterfly}
\end{figure}

The interaction of a Rydberg atom with a neutral atom located inside the classical Rydberg orbit can 
be modeled by means of the quasi-free electron picture, developed by Fermi~\cite{Fermi} in order to 
explain the line shift of Rydberg atoms in a buffer gas~\cite{Amaldi}. The original idea of Fermi was to
describe the electron-atom interaction with a delta-function potential proportional to the 
scattering length, the so-called Fermi pseudopotential, which in atomic units is given by
\begin{equation}
V_{\rm{Fermi}}( \vec{r},\vec{R} )=2\pi \textit{a}[k(R)]\delta^{(3)}(\vec{r}-\vec{R}),
\label{eq:Pseudopotential1}
\end{equation} 
where $\vec{r}$ denotes the electron position and $\vec{R}$ the perturber position, with respect to the Rydberg ionic core, and 
$a[k(R)]$ represents the energy dependent electron-perturber scattering length. For a given electron energy 
$\epsilon=-1/(2n^{*2})$, where $n^*$ is the effective principal quantum number, the electron-atom collision energy is $k^{2}(R)/2=\epsilon+1/R$. 

The Fermi pseudopotential turns out to be adequate for low-energy electrons, i.e., 
highly excited Rydberg states. However, in some systems the electron can be 
momentarily trapped due to the existence of shape resonances, thus 
Eq.~(\ref{eq:Pseudopotential1}) needs to be extended. Here, we study the case
 of \rubidium{}, which is known to have a shape resonance in the
\textsuperscript{3}P\textsuperscript{0} symmetry for \eRb{} scattering at
 \unit{0.02}{eV}~\cite{Fabrikant1986}. The method developed by Omont~\cite{Omont1977} is
 employed to generalize the Fermi pseudopotential including the 
 \textit{p}-wave shape resonance. Omont's approach is based on 
 approximating the zero-range pseudopotential with an \textit{l}-expansion 
 of the \textit{R}-matrix.  For the system at hand, only \textit{s}-wave and 
 \textit{p}-wave partial waves are necessary to characterize accurately the
  Rydberg-perturber interaction energy landscape,
\begin{eqnarray}
V_{\rm{Ryd-Perturb}}( \vec{r},\vec{R} ) = V_{\rm{Fermi}}( \vec{r},\vec{R} )- \nonumber \\
\frac{6\pi \tan{\left(\delta^{P}[k(R)]\right)}}{k(R)^{3}}\delta^{(3)}(\vec{r}-\vec{R})
\overleftarrow{\nabla}_{r}\cdot \overrightarrow{\nabla}_{r},
\label{eq:Omont}
\end{eqnarray} 
where $\delta^{P}[k(R)]$ stands for the triplet \textit{p}-wave scattering phase shift of \eRb{}. The arrow indicates that this should be read as an operator on the wavefunction, acting in the indicated direction.

Here, degenerate perturbation theory has been applied to calculate the Born-Oppenheimer PECs for the electron-atom interaction, as shown in Fig.~\ref{fig:butterfly}. In particular for the representation of the Rydberg-perturber potential given by Eq.~(\ref{eq:Omont}), a hydrogen-like basis for high angular momentum states has been used, whereas the Whittaker functions~\cite{Abramowitz:1974} have been utilized for low angular momentum states, thus taking into account the quantum defects. In each diagonalization eight different \textit{n}-manifolds have been included as well as their respective angular momentum states; in particular two of them are above the target state and six are below, taking into account the correct energy ordering of the states due to the quantum defects. The numerical results have been tested against the Green's function formalism~\cite{Hamilton2002} and the PECs are consistent even at intermediate ion-neutral separations, where the \textit{nS} state crosses the shape resonance potential. To verify the accuracy of the PECs at long range the resulting binding energies were compared to accurate spectroscopy data. In addition, for lower \textit{nS} states, we carried out test calculations including the spin-orbit interaction between the Rydberg electron and the perturber, and found only small deviations from the present results. This is in agreement with a previous investigation by A.\ Khuskivadze et al.~\cite{Khuskivadze2002}.

\begin{figure}[th!] 
	\includegraphics[scale=0.9]{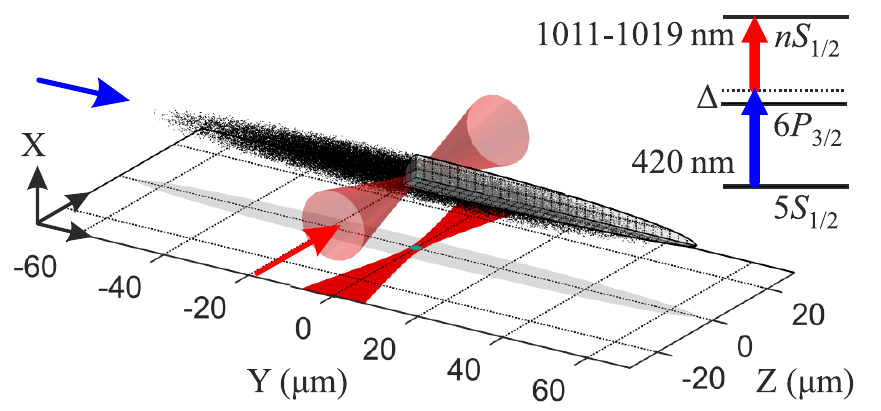}
	\caption{(Color online) Simplified schematic of a localized Rydberg excitation in our BEC drawn to scale. A focused infrared laser along the z-axis excites Rydberg atoms and determines the excitation volume inside the BEC together with a \unit{420}{nm} collimated beam along the y-axis.  The extent of the classical Rydberg electron orbital for the highest Rydberg state under investigation (111\textit{S}) is shown as tiny solid sphere (cyan) at the very center (filled circle on projection).  A simplified level scheme of our two-photon excitation is illustrated in the upper right corner.}
	\label{fig:excitation_scheme}
\end{figure}

\begin{figure}
\centering
	\includegraphics[scale = 1]{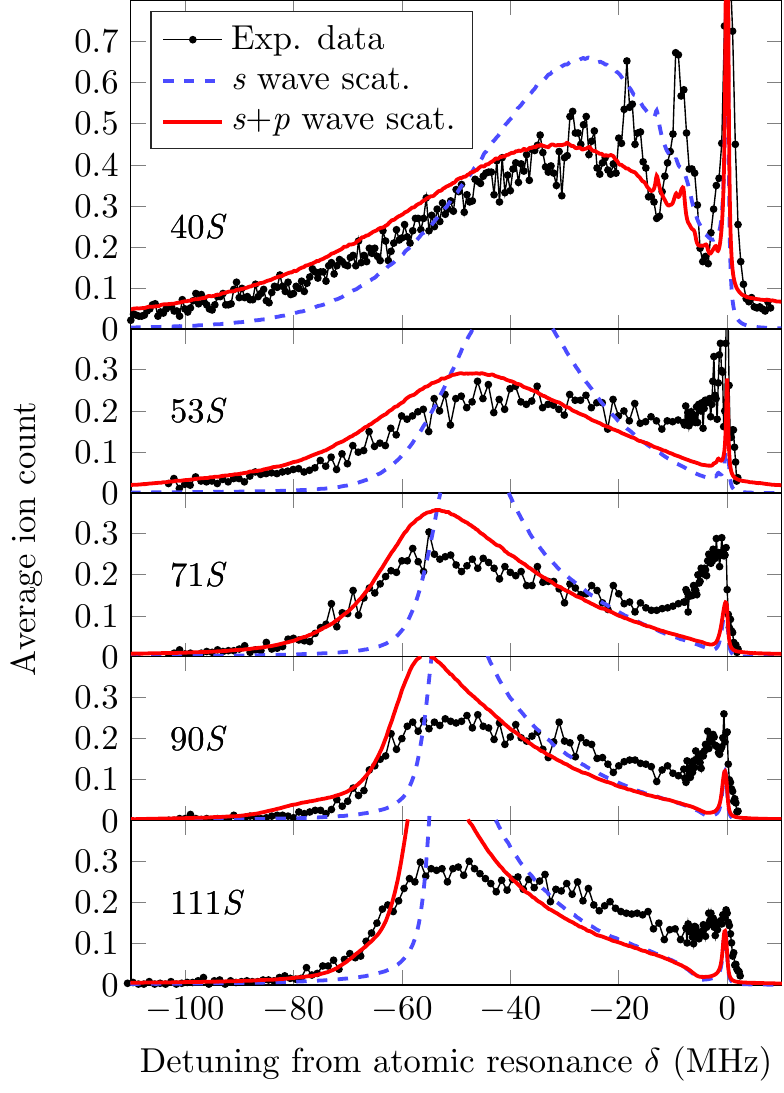}
	\caption{(Color online) Rydberg spectra in the BEC for different principal quantum numbers. The average ion count (black dots) is plotted versus the theory with \textit{s}-wave scattering components only (dashed blue) and the accurate PEC including also \textit{p}-wave components (red) of the \RydGS{} system, which includes the shape resonance. The zero detuning point is set by measuring the spectral position of the atomic line in a low density of 10\textsuperscript{12}\unit{}{cm\textsuperscript{-3}} and \unit{20}{\uK} thermal cloud where density and interaction effects can be neglected. The long tail to far red detuning and the signal at positive detuning taken at 40\textit{S} can only be explained by taking the \textit{p}-momentum components into account and therefore the shape resonance.}
	\label{fig:spectra}
\end{figure}

The experiment carried out and presented here to probe the PECs, utilize a single Rydberg atom in an almost pure spin polarized BEC of approximately 1.5$\times$10\textsuperscript{6} \rubidium{} atoms in the magnetically trapped ground state $\ket{\text{5}\textit{S}_{\text{1/2}}\text{,\,}\textit{F}\text{\,=\,2,\,}\textit{m}_{\textit{F}}\text{\,=\,2}}$. The trapping frequencies of the QUIC trap~\cite{Esslinger1998} are \trapfreq{r}{200} in the radial and \trapfreq{ax}{15} in the axial direction corresponding to Thomas-Fermi radii of \unit{5.0}{\um} by \unit{66}{\um} shown in Fig.~\ref{fig:excitation_scheme}.  The atom number and trap frequencies give rise to a peak density of the BEC of \density{5.5}{14}.  For the Rydberg excitation we apply a two-photon excitation scheme, where we couple the ground state to the $\ket{\textit{nS}_{\text{1/2}}\text{,\,}\textit{m}_{\textit{S}}\text{\,=\,1/2}}$ state for principal quantum numbers \textit{n} from 40 to 111, via the intermediate state 6\textit{P}\textsubscript{3/2}. For the lower transition we illuminate the cloud along the axial direction with pulsed \unit{420}{nm} light, \unit{2}{mm} in waist, to ensure a uniform Rabi frequency across the entire cloud. We use an intermediate state detuning of $\Delta$\,$=$\,\unit{80}{MHz}\,$\approx$\,56\,$\Gamma$ in order to keep absorption and heating of the BEC, due to scattering of the \unit{420}{nm} light, low. The upper transition to the Rydberg state is driven by an infrared laser with an \textit{n}-dependent wavelength between \unit{1011}{nm} and \unit{1019}{nm}, focused down to a (2.1$\pm$0.3)\,\um{} waist.  The \unit{420}{nm} excitation light is circularly polarized ($\sigma^+$), and the infrared light is set to have linear polarization along the x-axis perpendicular to the magnetic field in the y-direction. We pulse both excitation lasers simultaneously for \unit{2}{\us}, with a repetition rate of \unit{2}{kHz} to create \textit{nS} Rydberg atoms, which we subsequently electric field ionize within \unit{400}{ns} after the light fields turn off.  The field strength is set to be three times the ionization threshold field for the respective \textit{nS}-states, in order to also detect the Rydberg atoms that undergo inelastic state-changing collisions~\cite{Foltz1982,Rolfes1988}.  The ions are detected by a microchannel plate detector with an efficiency of 0.7.

For the data shown in Fig.~\ref{fig:spectra} the two-photon, single-atom Rabi frequency of \unit{250}{kHz} is kept constant for all the spectra by increasing the power of the infrared laser for higher Rydberg states, except for 40\textit{S} where the power of the infrared laser was twice as high. For the overall highest applied laser power, the trap depth of the time-averaged optical potential for this laser is about \unit{100}{nK}, which is less than the temperature of our BEC (\unit{300}{nK}).  The almost rectangular excitation pulse shape results in an excitation bandwidth of \unit{450}{kHz}~(FWHM). We choose a low single-atom Rabi frequency to minimize the probability of creating multiple Rydberg excitations within the BEC. Each point in the spectra represents 500 measurements, taken in 10 different clouds, to improve the signal-to-noise ratio. The mean atom number for each cloud within the 50 shots was 1.4$\times$10\textsuperscript{6}\,BEC and 0.3$\times$10\textsuperscript{6}\,thermal atoms with a temperature of \unit{300}{nK}.

The sharp peaks visible in the experimental data at small red detuning in the 40\textit{S} and 53\textit{S} spectra are due to the formation of Rydberg-ground state molecules \cite{Bendkowsky2009} and indicate the spectral resolution that we are sensitive to. The overall broad spectral feature is mainly signal from the Rydberg excitations in the BEC. Taking the peak density, $\rho$\,=\,\density{5.5}{14}, of our BEC into account, in combination with the mean field density shift introduced by Fermi~\cite{Fermi}, one expects a maximum in the signal around \unit{-55}{MHz}.  The spectroscopic data show, however, a pronounced \textit{n}-dependent line shift and broadening extending the signal to much larger red and blue detuning.  

\begin{figure}
\centering
	\includegraphics[scale=1]{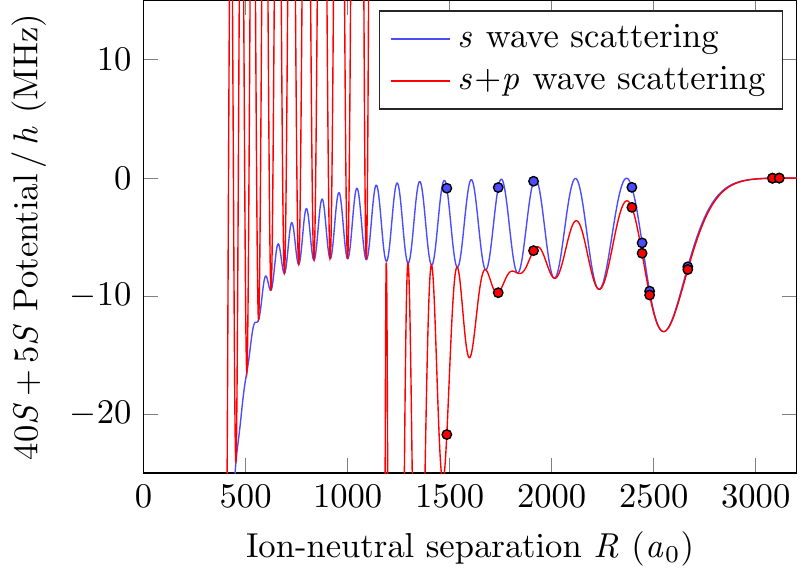}
	\caption{(Color online) Potential energy curves of the \RydSpecificGS{} with \textit{s}-wave (blue) momentum components of the Rydberg electron only and including \textit{p}-wave scattering (red). As an example particles (dots) are placed according to a Poissonian density distribution corresponding to a density of \density{5}{14} resulting in a shift of the resonance of \unit{-25.5}{MHz} for \textit{s}-wave scattering only (blue dots) and \unit{-64.2}{MHz} including also \textit{p}-wave scattering (red dots). The mean inter-particle separation at this density is \unit{1300}{\rbohr}.}
	\label{fig:comparison}
\end{figure}

In Fig.~\ref{fig:comparison} the accurate potential, which 
includes \textit{s}- and \textit{p}-wave scattering and therefore the shape resonance is compared to the potential that incorporates only \textit{s}-wave scattering. At our peak densities the mean inter-particle distance is \unit{1300}{a0}, which is comparable to the \textit{p}-wave resonance position relative to the ionic core of the Rydberg atom. At the time of the excitation of a \RydNGS{} state, where \textit{N} is the number of the neutral atom perturbers within the Rydberg orbit, there is a high probability of finding one atom close to crossing of the \textit{nS} state and shape resonance potential. It is apparent that atoms excited at the crossing have a strong impact on the line shape broadening as observed in the experimental and modeled spectra shown in Fig.~\ref{fig:spectra}. The position of the crossing of the shape resonance potential with the \textit{nS} state stays within \unit{1200}{\rbohr} for \textit{n}\,=\,40 and the asymptotic value of approximately \unit{2000}{\rbohr} for high \textit{n}, whereas the slope of the crossing scales as \textit{n}\textsuperscript{$*$-4} for the investigated principal quantum numbers. Therefore the line shape broadening is \textit{n}-dependent, and more severe for lower quantum numbers. 

To calculate the modeled spectra of Fig.~\ref{fig:spectra}, atoms are placed randomly at positions $\vecx_i$ according to the measured bimodal distribution in the experiment. The BEC is assumed to have a perfect Thomas-Fermi distribution as shown in Fig.~\ref{fig:excitation_scheme}, whereas the thermal cloud has a Gaussian profile according to the measured trap frequencies and temperature. All atoms are treated as spatially uncorrelated assuming T\,=\,0. For every atom in the sample the shift, $\delta_i$, from the atomic resonance is calculated by
\begin{equation}
\delta_i = \sum_{j \ne i} V_i(\left|\vecx_j - \vecx_i\right|),
\end{equation}
using the atoms located at $\vecx_j$ surrounding the Rydberg atom $\vecx_i$.  These atoms are treated as point-like particles in the \RydGS{} potential $V$ as shown in Fig.~\ref{fig:comparison}.
The spectrum $S(\delta)$ is finally calculated by summing over all Lorentzian contributions of each atom, taking into account its shift $\delta_i$, the excitation bandwidth $\Gamma$, and the spatially varying excitation laser intensity $I_i=I(\vecx_i)$ relating to the incoherent excitation rate in our system. Rydberg blockade effects~\cite{Saffman} are neglected since the mean ion count rate is well below one in the area of interest. Each simulated spectrum is averaged over several atom configurations and normalized with a factor $a$ to have the same $\int{}\!S(\delta)d\delta$ compared to the experimental data.
\begin{equation}
S(\delta) = a \sum_i \frac{1}{\pi} \frac{\left(\frac{1}{2} \Gamma\right)^2}{\delta_i^2+\left(\frac{1}{2}{\Gamma}\right)^2} \times I_i
\end{equation}
 There are no free parameters used for the simulation.

Comparing the experimental data and the simulated spectra of Fig.~\ref{fig:spectra}, the significance of the \textit{p}-wave scattering, and therefore the crossing of the \textit{nS} state with the shape resonance potential, is apparent. Only by including the shape resonance, the long tail of the spectra towards red detuning and the signal at blue detuning are modeled correctly. The signal at blue detuning comes from atoms excited on the upper branch of the \textit{nS} and shape resonance crossing (see Fig.~\ref{fig:comparison} for \unit{500}{\rbohr}\,$<$\,\textit{r}\,$<$\,\unit{1200}{\rbohr}). The Rydberg molecular lines, especially visible in the 40\textit{S} and 53\textit{S} spectra, cannot be reproduced by evaluating the PECs with classical point-like particles. For higher principal quantum numbers, the theory spectra show too much signal between -70 and \unit{-40}{MHz}. The spectra taken show that Rydberg atoms with a detuning in this range are not shifted as much as the theory predicts, leading to an underestimated theory signal between -40 and \unit{0}{MHz}.  There are several simplifications in the model, which could lead to these discrepancies at intermediate detunings. All atoms are treated as point-like particles and their wave character, relevant at these low temperatures, is neglected. In addition the atom distribution is assumed to be bimodal and uncorrelated, i.e. bunching is not taken into account for finite temperatures \cite{Schmidt2015}.  It was verified that spin-orbit coupling does not alter the spectra modeled for low \textit{n} states, but it could play a role for the modeled spectra for higher \textit{n}. It could also be that the pairwise PECs are not valid any more if a neutral atom is excited within the crossing. In the model we neglect any resulting backaction on the electron density and therefore the PECs. Since the position of the crossing changes with \textit{n}, the mean number of atoms close to the resonance changes also from 0.3 at \textit{n}\,=\,40 to 1.1 at \textit{n}\,=\,111 for the peak densities in the BEC. 

Our model simulates the microscopic electron-atom interactions inside the Rydberg orbit.  Such a microscopic model, in contrast to the models employed in the previous theory work~\cite{Fabrikant1986, Borodin1990, Borodin1991, Henry2002}, lends itself to modeling the dynamics of the neutral atoms within the Rydberg atom.  We expect that at the cold temperatures realized in this experiment (\unit{\textless\,1}{\uK}),  the neutral atom perturbers will evolve in the PECs presented here, leading to a limitation of the collision lifetime of Rydberg atoms excited in dense media~\cite{LossesNature,Niederpruen2015}.  The evolution of the perturbing neutral atom wavefunction in the shape resonance potential is a new aspect to utilizing Rydberg excitations in dense, cold media. These dynamics must be considered for future proposals relying on utilizing a Rydberg atom in a BEC~\cite{Wavefunctionimaging, Wang2015, Mukherjee2015}.  We have shown that the line broadening of the Rydberg levels in an ultracold and dense media is a direct manifestation of the underlying Rydberg-perturber PECs, by accounting only for elastic \eRb{} scattering.  In this range of \textit{n}-values, inelastic collisions do not, seemingly, lead to significant line broadening.  

The results presented are based on previous understanding of scattering events starting with Fermi~\cite{Fermi} and leading to a mean-field energy shift~\cite{LossesNature,Gaj2014}. But to explain the high resolution spectra taken here in a BEC with very high densities, the potential energy curves including \textit{s}- and \textit{p}-wave scattering has to be taken into account and therefore the shape resonance in Rb. The present model indicates that in systems free of \textit{p}-wave resonances, for instance in Sr, the many-body approach of the mean-field energy shift will be a suitable tool for describing the physics of the interaction between the Rydberg and the BEC. Finally, it is shown that such Rydberg spectroscopy in an ultracold and dense environment, is a tool for testing potential energy curves for electron-atom interactions and the scattering phase shifts.

\begin{acknowledgments}
We acknowledge support from Deutsche Forschungsgemeinschaft (DFG) within the 
SFB/TRR21 and the project PF 381/13-1. Parts of this work was also founded by
ERC under contract number 267100. S.H.\ acknowledges support from DFG through 
the project HO 4787/1-1, H.N. acknowledges support from the Deutsche Studienstiftung and M.S. acknowledges support from the Carl Zeiss Foundation. 
This work has been supported by NSF under grand number PHY-130690.
\end{acknowledgments}

\bibliography{literature}

\end{document}